\documentclass[prl,groupedaddress,showpacs,nofootinbib,twocolumn]{revtex4}
\usepackage[pdftex]{graphicx}
\usepackage{amssymb}
\newcommand{\ped}[1]{\ensuremath{_{\rm #1}}}

\begin{document}

\title{Three-band s$\pm$ Eliashberg theory and the superconducting gaps of iron pnictides}

\author{G.A. Ummarino} \email{E-mail:giovanni.ummarino@infm.polito.it}
\author{M. Tortello}
\author{D. Daghero}
\author{R.S. Gonnelli}

\affiliation{Dipartimento di Fisica and CNISM, Politecnico di
Torino, Corso Duca degli Abruzzi 24, 10129 Torino, Italy}

\begin{abstract}
The experimental critical temperatures and gap values of the
superconducting pnictides of both the 1111 and 122 families can be
simultaneously reproduced within the Eliashberg theory by using a
three-band model where the dominant role is played by interband
interactions and the order parameter undergoes a sign reversal
between hole and electron bands ($s \pm$-wave symmetry). High
values of the electron-boson coupling constants and small typical
boson energies (in agreement with experiments) are necessary to
obtain the values of all the gaps and to correctly reproduce their
temperature dependence.

\end{abstract}


\maketitle

The recently discovered Fe-based pnictide superconductors
\cite{Kamihara_La, Ren_Sm, Rotter_122} have aroused great interest
in the scientific community. They have indeed shown that high T\ped
c superconductivity does not uniquely belong to cuprates but can
take place in Cu-free systems as well. Nevertheless, as in cuprates,
superconductivity occurs upon charge doping of a magnetic parent
compound above a certain critical value. However, important
differences exist: the parent compound in cuprates is a Mott
insulator with localized charge carriers and a strong Coulomb
repulsion between electrons; in the pnictides, on the other hand, it
is a bad metal and shows a tetragonal to orthorhombic structural
transition below $\approx$ 140 K, followed by an antiferromagnetic
(AF) spin-density-wave (SDW) order \cite{de la Cruz}. Charge doping
gives rise to superconductivity and, at the same time, inhibits the
occurrence of both the static magnetic order and the structural
transition. The Fermi surface consists of two or three hole-like
sheets around $\Gamma$ and two electron-like sheets around $M$. Up
to now, the most intensively studied systems are the 1111 compounds,
ReFeAsO$\ped{1-x}$F$\ped x$ (Re = La, Sm, Nd, Pr, etc.) and
especially the 122 ones, hole- or electron-doped AFe$\ped 2$As$\ped
2$ (A = Ba, Sr, Ca). The huge amount of experimental work already
done in 122 compounds is due to the availability of rather big
high-quality single crystals.

Most of the present research effort is spent clarifying the
microscopic pairing mechanism responsible for superconductivity. The
conventional phonon-mediated coupling mechanism cannot explain the
observed high T\ped c within standard Migdal-Eliashberg theory and
the inclusion of multiband effects increases T\ped c only marginally
\cite{Boeri_PhysC_SI}. On the other hand, the magnetic nature of the
parent compound seems to favor a magnetic origin of
superconductivity and a coupling mechanism based on nesting-related
AF spin fluctuations has been proposed \cite{Mazin_spm}. It predicts
an interband sign reversal of the order parameter between different
sheets of the Fermi surface ($s \pm$ symmetry). The number,
amplitude and symmetry of the superconducting energy gaps are indeed
fundamental physical quantities that any microscopic model of
superconductivity has to account for. Experiments with powerful
techniques such as ARPES, point-contact spectroscopy, STM etc., have
been carried out to study the superconducting gaps in pnictides (for
a review see \cite{PhysC_Fe}). Although results are sometimes in
disagreement with each other, a multi-gap scenario is emerging with
evidence for rather high gap ratios, $\Delta_1 / \Delta_2 \approx
2-3$ \cite{PhysC_Fe}. A two-band BCS model cannot account either for
the amplitude of the experimental gaps and for their ratio.
Three-band BCS models have been investigated \cite{Mazin_PhysC_SI,
Benfatto, Kuchinskii} which can reproduce the experimental gap ratio
but not the exact experimental gap values. In this regard a reliable
study has to be carried out within the framework of the Eliashberg
theory for strong coupling superconductors \cite{Eliashberg}, due to
the possible high values of the coupling constants necessary to
explain the experimental data.

By using this strong-coupling approach, we show here that the
superconducting iron pnictides represent a case of dominant negative
interband-channel superconductivity ($s \pm$-wave symmetry) with
high values of the electron-boson coupling constants and small
typical boson energies. Furthermore we prove that a small
contribution of intraband coupling does not affect significantly the
obtained results. The model is compared with the results of two
representative experiments in 122 \cite{Ding_ARPES_BaKFeAs} and 1111
\cite{Daghero_Sm} compounds proving to be able to reproduce fairly
well the values and the whole temperature dependence of the
superconducting energy gaps.

\begin{figure}[b]
\begin{center}
\includegraphics[keepaspectratio, width=0.2\textwidth]{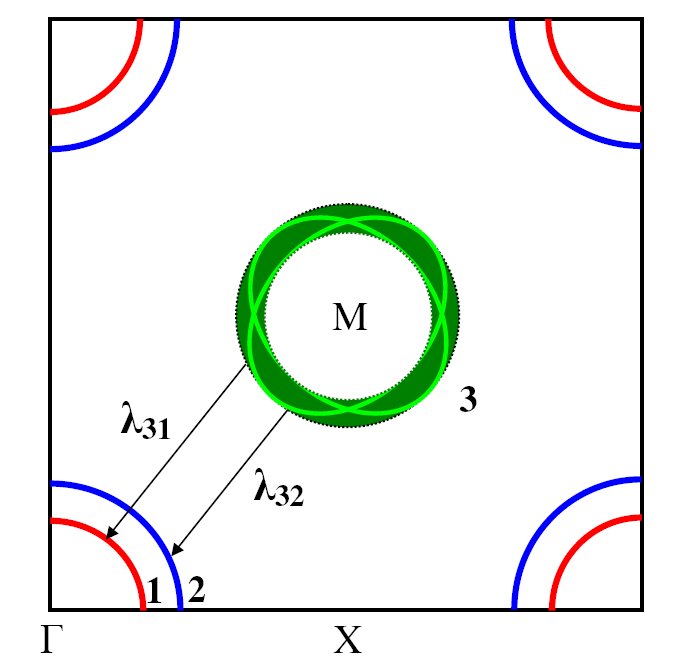}
\vspace{-4mm} \caption{Schematic drawing of the multiband model used
in this work. The two hole bands (1 and 2) are centered around the
$\Gamma$ point, while the equivalent electron band (3) around the
$M$ point of the reduced Brillouin zone.}\label{Fig1}
\end{center}
\vspace{-4mm}
\end{figure}

As a starting point we can model the electronic structure of
pnictides by using a three-band model (Fig.\ref{Fig1}) with two hole
bands (1 and 2) and one equivalent electron band (3)
\cite{Mazin_PhysC_SI}. The $s$-wave order parameters of the hole
bands have opposite sign with respect to that of the electron one
\cite{Mazin_spm}. Intraband coupling could be provided by phonons
while interband coupling by antiferromagnetic spin fluctuations. In
a one-band system spin fluctuations (SF) are always pair breaking
but in a multiband one the interband term can contribute to increase
the critical temperature. Indeed, in the multiband Eliashberg
equations (EE) the SF term in the intraband channel has positive
sign for the renormalization functions $Z_{i}$ and negative sign for
the superconducting order parameters $\Delta_{i}$ thus leading to a
strong reduction of T$_c$. However, if we consider negative
interband contributions in the $\Delta_i$ equations, the final
result can be an increase of the critical temperature
\cite{Ummaimp}.

Let us consider the generalization of the Eliashberg theory
\cite{Eliashberg} for multiband systems, that has already been
used with success to study the MgB$_{2}$ superconductor
\cite{Carbinicol}. To obtain the gaps and the critical temperature
within the s-wave, three-band Eliashberg equations one has to
solve six coupled integral equations for the gaps
$\Delta_{i}(i\omega_{n})$ and the renormalization functions
$Z_{i}(i\omega_{n})$, where $i$ is a band index that ranges
between $1$ and $3$ (see Fig.\ref{Fig1}) and $\omega_{n}$ are the
Matsubara frequencies. For completeness we included in the
equations the non-magnetic and magnetic impurity scattering rates
in the Born approximation, $\Gamma^{N}_{ij}$ and
$\Gamma^{M}_{ij}$:
\begin{eqnarray}
\omega_{n}Z_{i}(i\omega_{n})=\omega_{n}+\sum_{j}(\Gamma^{N}_{ij}+\Gamma^{M}_{ij})N^{Z}_{j}(i\omega_{n})+\label{eq:EE1}\\\nonumber
\pi
T\sum_{m,j}\Lambda^{Z}_{ij}(i\omega_{n},i\omega_{m})N^{Z}_{j}(i\omega_{m})
\end{eqnarray}
\begin{eqnarray}
Z_{i}(i\omega_{n})\Delta_{i}(i\omega_{n})=\sum_{j}(\Gamma^{N}_{ij}-\Gamma^{M}_{ij})N^{\Delta}_{j}(i\omega_{n})+\label{eq:EE2}\\\nonumber
\pi
T\sum_{m,j}[\Lambda^{\Delta}_{ij}(i\omega_{n},i\omega_{m})\nonumber
-\mu^{*}_{ij}(\omega_{c})]\theta(\omega_{c}-|\omega_{m}|)N^{\Delta}_{j}(i\omega_{m})\nonumber
\end{eqnarray}
where
$\Lambda^{Z}_{ij}(i\omega_{n},i\omega_{m})=\Lambda^{ph}_{ij}(i\omega_{n},i\omega_{m})+\Lambda^{sp}_{ij}(i\omega_{n},i\omega_{m})$,
$\Lambda^{\Delta}_{ij}(i\omega_{n},i\omega_{m})=\Lambda^{ph}_{ij}(i\omega_{n},i\omega_{m})-\Lambda^{sp}_{ij}(i\omega_{n},i\omega_{m})$.
$\theta$ is the Heaviside function and $\omega_{c}$ is a cut-off
energy. In particular,
$\Lambda^{ph,sp}_{ij}(i\omega_{n},i\omega_{m})=\int_{0}^{+\infty}d\Omega
\alpha^{2}_{ij}F^{ph,sp}(\Omega)/[(\omega_{n}-\omega_{m})^{2}+\Omega^{2}]$,
where $ph$ means \textquotedblleft phonon\textquotedblright$\;$
and $sp$ \textquotedblleft spin fluctuations\textquotedblright.
Finally, $N^{\Delta}_{j}(i\omega_{m})=\Delta_{j}(i\omega_{m})/
{\sqrt{\omega^{2}_{m}+\Delta^{2}_{j}(i\omega_{m})}}$ and
$N^{Z}_{j}(i\omega_{m})=\omega_{m}/{\sqrt{\omega^{2}_{m}+\Delta^{2}_{j}(i\omega_{m})}}$.

In principle the solution of the three-band EE shown in
eqs.\ref{eq:EE1} and \ref{eq:EE2} requires a huge number of input
parameters: i) nine electron-phonon spectral functions,
$\alpha^{2}_{ij}F^{ph}(\Omega)$; ii) nine electron-SF spectral
functions, $\alpha^{2}_{ij}F^{sp}(\Omega)$; iii) nine elements of
the Coulomb pseudopotential matrix, $\mu^{*}_{ij}(\omega\ped{c})$;
iv) nine non-magnetic ($\Gamma^{N}_{ij}$) and nine paramagnetic
($\Gamma^{M}_{ij}$) impurity scattering rates.\\
It is obvious that a practical solution of these equations requires
a drastic reduction in the number of free parameters of the model.
On the other hand, from the work of Mazin et al.
\cite{Mazin_PhysC_SI} we know that: i)
$\lambda^{ph}_{ii}>>\lambda^{ph}_{ij}\approx0$ i.e. phonons mainly
provide intraband coupling but the total electron-phonon coupling
constant $\Sigma_{i}\lambda^{ph}_{ii}$ should be very small
\cite{Boeri_PhysC_SI}, ii)
$\lambda^{sp}_{ij}>>\lambda^{sp}_{ii}\approx0$, i.e. SF mainly
provide interband coupling. We include these features in the most
simple three-band model by posing:
$\lambda^{ph}_{ii}=\lambda^{ph}_{ij}=0$, $\lambda^{sp}_{ii}=0$ and
$\mu^{*}_{ii}(\omega\ped{c})=\mu^{*}_{ij}(\omega\ped{c})=0$. In
addition, we set $\Gamma^{N}_{ij} = \Gamma^{M}_{ij} = 0$ in eqs. (1)
and (2).

Under these approximations, the electron-boson coupling-constant
matrix is then \cite{Mazin_PhysC_SI}:

\begin{displaymath}
\left (
\begin{array}{ccc}
  0 & 0 & \lambda_{31}\nu_{1} \\
  0 & 0 & \lambda_{32}\nu_{2} \\
  \lambda_{31} & \lambda_{32} & 0 \\
\end{array}
\right )
\end{displaymath}

where $\nu_{1}=N_{1}(0)/N_{3}(0)$, $\nu_{2}=N_{2}(0)/N_{3}(0)$ and
$N_{i}(0)$ is the normal density of states at the Fermi level for
the $i$-band ($i=1, 2, 3$ according to Fig.1).

We initially solved the EE on the imaginary axis to calculate the
critical temperature and, by means of the technique of the Pad\`{e}
approximants, to obtain the low-temperature value of the gaps. In
presence of a strong coupling interaction or of impurities, however,
the value of $\Delta_{i}(i\omega_{n=0})$ obtained by solving the
imaginary-axis EE can be very different from the value of
$\Delta_{i}$ obtained from the real-axis EE \cite{Ummastrong}.
Therefore, in order to determine the exact temperature dependence of
the gaps, we then solved the three-band EE in the real-axis
formulation.
\begin{figure}[t]
\begin{center}
\includegraphics[keepaspectratio, width=0.45\textwidth]{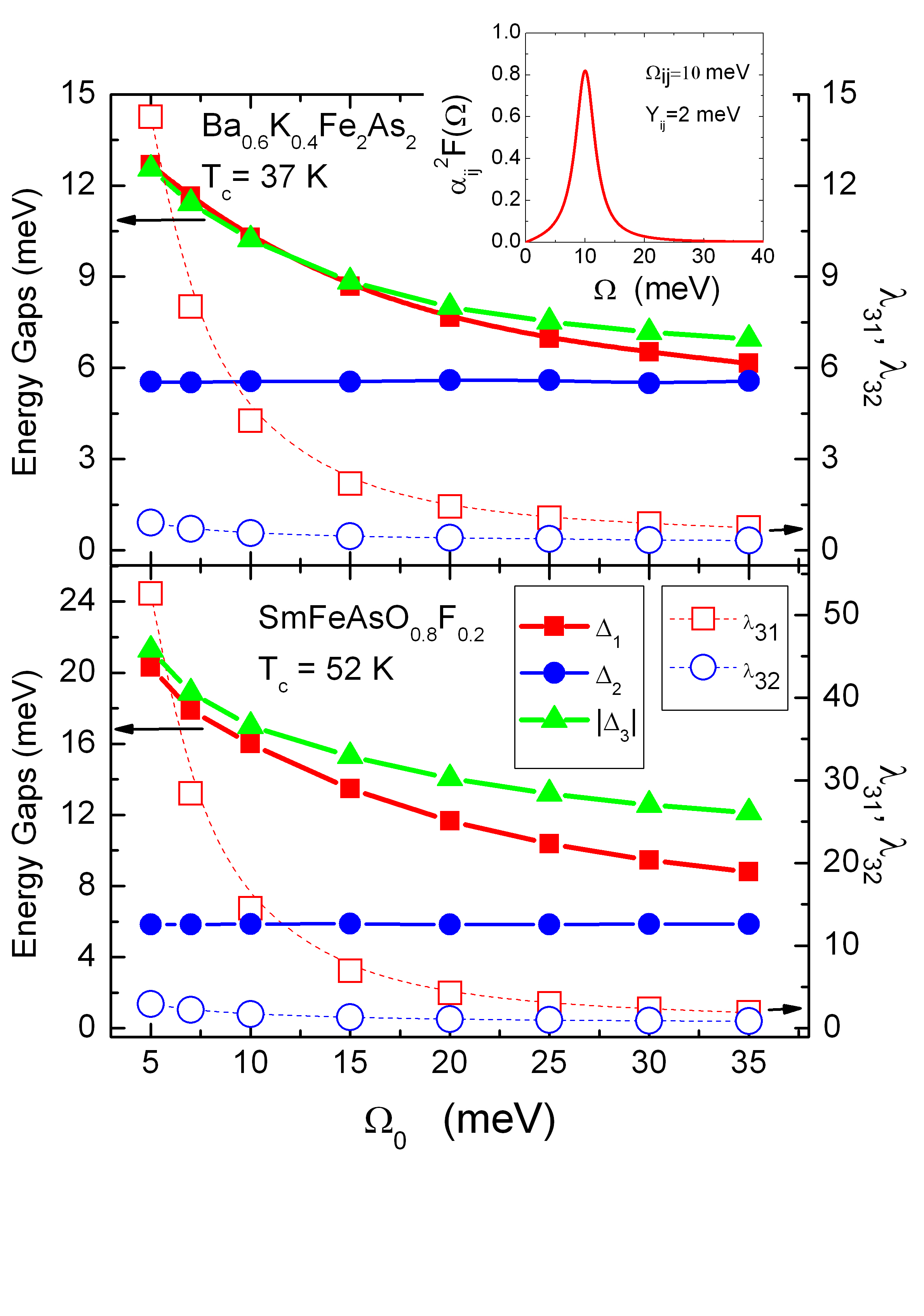}
\vspace{-18mm} \caption{Full symbols, left axis: Calculated gap
values at $T$=2 K for Ba$_{0.6}$K$_{0.4}$Fe$_2$As$_2$ (upper panel)
and SmFeAsO$_{0.8}$F$_{0.2}$ (lower panel) as function of typical
boson energy $\Omega_0$. Open symbols, right axis: Electron-boson
coupling constants, $\lambda_{31}$ and $\lambda_{32}$ as function of
$\Omega_0$. The inset shows the spectral function used in this model
in the case $\Omega_{ij} = 10$ meV.}\label{Fig2}
\end{center}
\vspace{-8mm}
\end{figure}

We tried to reproduce the critical temperature and the gap values in
two representative cases: i) the 122 compound
Ba$_{0.6}$K$_{0.4}$Fe$_{2}$As$_{2}$ with $T_{c}=37$ K where ARPES
measurements gave $\Delta_{1}(0) \approx 12.1$ meV, $\Delta_{2}(0)
\approx 5.5$ meV and $\Delta_{3}(0) \approx 12.8$ meV
\cite{Ding_ARPES_BaKFeAs}; ii) the 1111 compound
SmFeAsO$_{0.8}$F$_{0.2}$ with $T_{c}=52$ K where from point-contact
spectroscopy measurements we obtained $\Delta_{1}(0)=18 \pm 3$ meV
and $\Delta_{2}(0)=6.15 \pm 0.45$ meV \cite{Daghero_Sm}.

Inelastic neutron-scattering experiments suggest that the typical
boson energy possibly responsible for superconductivity ranges
roughly between 10 and 30 meV \cite{neutroni}. In our numerical
simulations we used spectral functions with Lorentzian shape, i.e.
$\alpha^{2}_{ij}F(\Omega)=C_{ij}[L(\Omega+\Omega_{ij},Y_{ij})-L(\Omega-\Omega_{ij},Y_{ij})]$
where
$L(\Omega\pm\Omega_{ij},Y_{ij})=[(\Omega\pm\Omega_{ij})^{2}+(Y_{ij})^{2}]^{-1}$,
$C_{ij}$ are the normalization constants necessary to obtain the
proper values of $\lambda_{ij}$, while $\Omega_{ij}$ and $Y_{ij}$
are the peak energies and half-widths, respectively. In all our
calculations we always set $\Omega_{ij}=\Omega_{0}$, with $\Omega_0$
ranging between 5 and 35 meV and $Y_{ij}=2$ meV. The cut-off energy
is $\omega_{c}=12 \cdot\Omega_{0}$ and the maximum quasiparticle
energy is $\omega_{max}=16 \cdot\Omega_{0}$.

In the 122 case ($T_{c}=37$ K) we know that $\nu_{1}=1$ and
$\nu_{2}=2$ \cite{Mazin_PhysC_SI} while in the 1111 case ($T_{c}=52$
K) we have $\nu_{1}=0.4$ and $\nu_{2}=0.5$ \cite{Mazin4}. Once the
energy of the boson peak, $\Omega_0$ is set, only two free
parameters are left in the model: $\lambda_{31}$ and $\lambda_{32}$.

By properly selecting the values of these parameters it is
relatively easy to obtain the experimental values of the critical
temperature and of the small gap. It is more difficult to reproduce
the values of the large gaps of band 1 and 3 since, due to the high
$2\Delta_{1,3} / k_B T_c$ ratio (of the order of 8-9), high values
of the coupling constants and small boson energies are required.
Fig.\ref{Fig2} shows the values of the calculated gaps (full
symbols, left axis) as function of the boson peak energy,
$\Omega_0$. The corresponding values of $\lambda_{31}$ and
$\lambda_{32}$, chosen in order to reproduce the values of T$_c$ and
of the small gap, $\Delta_2$, are also shown in the figure (open
symbols, right axis). In both materials, only when $\Omega_{0}\leq
10$ meV the value of the large gaps correspond to the experimental
data. Indeed, when $\Omega_0$ increases, the values of $\Delta_1$
and $\Delta_3$ strongly decrease. As a consequence, a rather small
energy of the boson peak together with a very strong coupling
(particularly in the 3-1 channel) is needed in order to obtain the
experimental $T_c$ and the correct gap values. In this regard, it is
worth noticing that the absolute values of the large gaps
\emph{cannot} be reproduced in a interband-only, two-band Eliashberg
model \cite{dolghi3} as well as within a three-band BCS model. In
the latter case it is only possible to obtain a ratio of the gaps
close to the experimental one \cite{Kuchinskii,Benfatto}.

\begin{figure}[t]
\begin{center}
\includegraphics[keepaspectratio, width=0.4\textwidth]{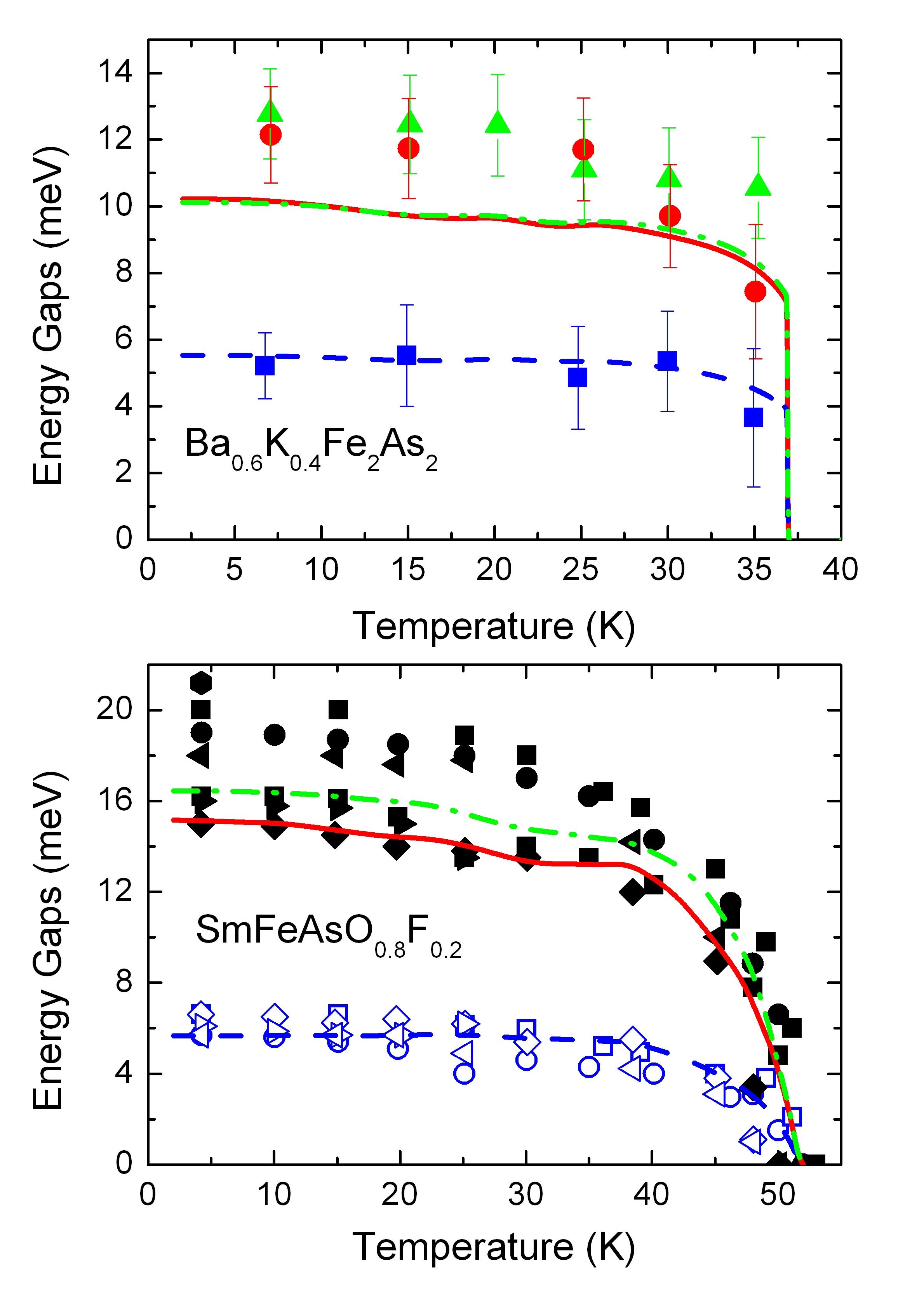}
\vspace{-5mm} \caption{Calculated temperature dependence of the gaps
for Ba$_{0.6}$K$_{0.4}$Fe$_2$As$_2$ ($T_{c}=37$ K, upper panel) and
for SmFeAsO$_{0.8}$F$_{0.2}$ ($T_{c}=52$ K, lower panel):
$\Delta_1(T)$ (red solid line), $\Delta_2(T)$ (blue dashed line) and
$\Delta_3(T)$ (green dash-dot line). Symbols are experimental data
from ref. \cite{Ding_ARPES_BaKFeAs} (upper panel) and ref.
\cite{Daghero_Sm} (lower panel).}\label{Fig3}
\end{center}
\vspace{-8mm}
\end{figure}

\begin{table}[b]
\begin{tabular}{|c|c|c|}
\hline
   $\lambda_{31}=3.866$  & $\lambda_{32}=0.471$ & $\mu^{*}_{ij}=0$, $\lambda_{ii}=0.4$\\
   \hline
  $\Delta_{1}=10.30$ meV & $\Delta_{2}= 5.62$ meV & $|\Delta_{3}|=10.24$ meV\\
\hline
\end{tabular}
\caption{The effect of a small contribution of intraband coupling,
$\lambda_{ii}=0.4$ for the case of Ba$_{0.6}$K$_{0.4}$Fe$_2$As$_2$
at $T=2$ K.}\label{Table1}
\end{table}

Another important result of the model is the temperature
dependence of the gaps. Figure \ref{Fig3} shows this dependence
for the experimental gaps (symbols) together with the theoretical
$\Delta_i(T)$ curves obtained by the three-band Eliashberg model
(lines) for Ba$_{0.6}$K$_{0.4}$Fe$_2$As$_2$ (upper panel) and
SmFeAsO$_{0.8}$F$_{0.2}$ (lower panel). The parameters used for
the 122 compound are $\Omega_{0}=10$ meV, $\lambda_{31}=4.267$ and
$\lambda_{32}=0.569$; for the 1111 compound we used
$\Omega_{0}=10$ mev, $\lambda_{31}=14.520$ and
$\lambda_{32}=1.708$. The experimental temperature dependence of
the gaps shown in the upper panel is rather unusual with the gaps
slightly decreasing with increasing temperature until they
suddenly drop close to $T_c$. The theory reproduces very well this
behavior, which is possible only in a very strong coupling regime
\cite{Ummastrong}. The different temperature dependence observed
in the lower panel of Fig. \ref{Fig3} results from a complex
non-linear dependence of $\Delta_i$ vs. $T$ curves on
$\lambda_{31}$. Further details will be given in a forthcoming
paper.

We also tested the effect into the model of a small intraband
coupling (possibly of phonon origin). In the case of
Ba$_{0.6}$K$_{0.4}$Fe$_2$As$_2$ we used $\lambda_{ii}=0.4$ since
we know indeed that this coupling cannot be very high
\cite{Boeri_PhysC_SI}. It might be thought that this term can
sensibly contribute to increase the gap values but, as can be seen
in Table \ref{Table1}, this is not the case: the gap values only
show a slight increase (of the order of 1\%).
\begin{table}[t]
\begin{tabular}{|c|c|c|}
\hline
  $\lambda_{31}=2.730$ &  $\lambda_{32}=0.758$ & $\mu^{*}_{ij}=0.1$, $\lambda_{ii}=0.4$ \\
\hline
  $\Delta_{1}=7.49$ meV & $\Delta_{2}= 5.72$ meV & $|\Delta_{3}|=7.98$ meV\\
  \hline
\end{tabular}
\caption{The effect of the Coulomb interaction, $\mu^{*}_{ij}$ for
the case shown in Table \ref{Table1}.}\label{Table2}
\vspace{-5mm}
\end{table}

The effect of Coulomb interaction was also investigated for the case
shown in Table \ref{Table1} where a weak intraband coupling is
included. We chose $\mu^{*}_{ij}=0.1$ and, as expected, we found
that the intraband Coulomb pseudopotential has a negligible effect
while the interband one \cite{Ummaimp} strongly contributes to raise
$T_{c}$ and reduces in a considerable way the value of
$\lambda_{31}$: in this case, as shown in Table \ref{Table2}, it is
only possible to obtain the correct value of the small gap. As a
consequence, this result seems to exclude a strong interband Coulomb
interaction in these compounds.

\begin{figure}[t]
\begin{center}
\includegraphics[keepaspectratio, width=0.35\textwidth]{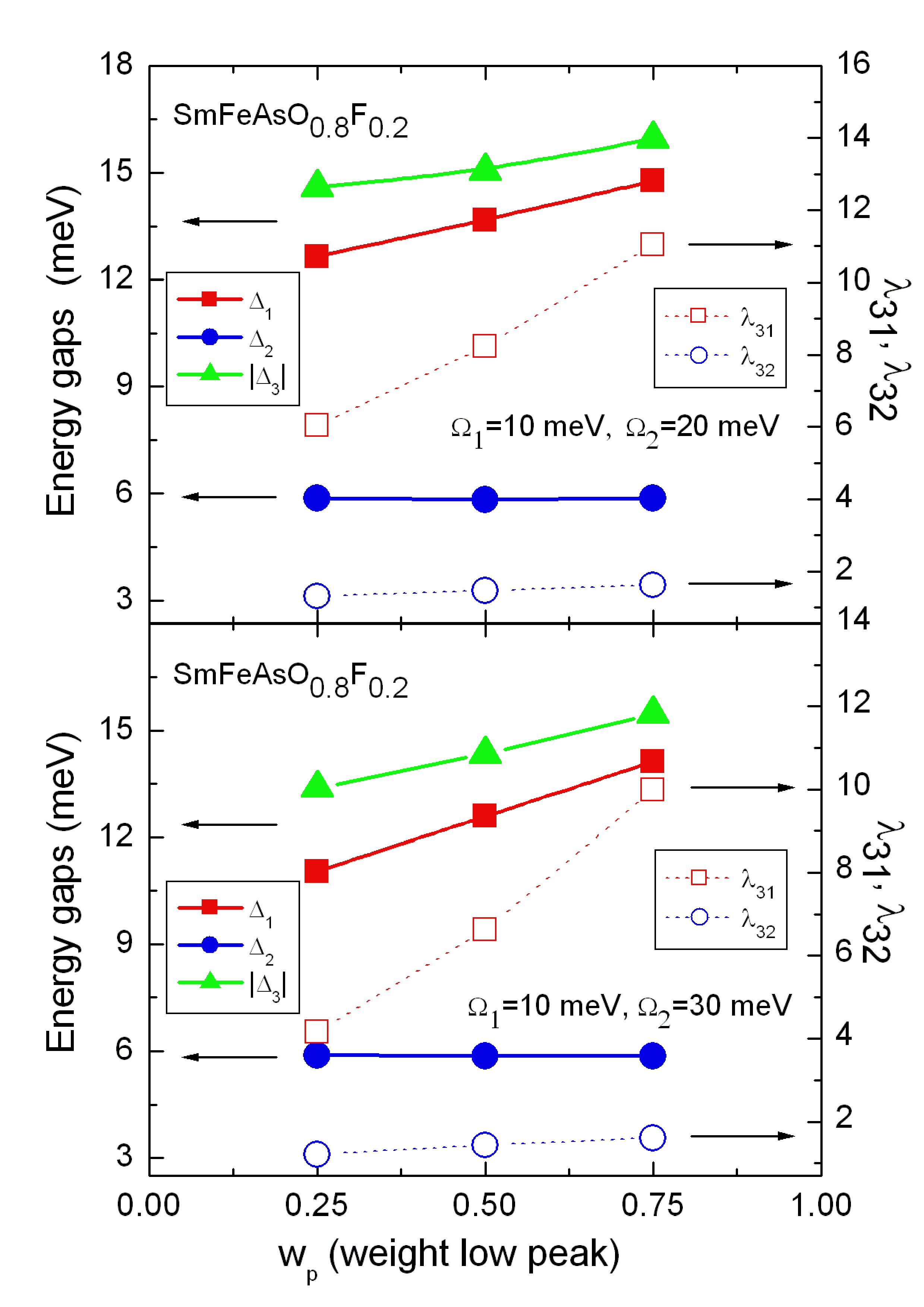}
\vspace{-5mm} \caption{The calculated gaps and electron-boson
coupling constants $\lambda_{31}$ and $\lambda_{32}$ in the 1111
case as function of the weight $w_{p}$ of the low-energy peak
$\Omega_{1}$. $\Omega_{2}=20$ meV (upper panel) and 30 meV (lower
panel).}\label{Fig4}
\end{center}
\vspace{-11mm}
\end{figure}

Finally we have also examined, for SmFeAsO$_{0.8}$F$_{0.2}$, the
case of a spectral function with two peaks at energies $\Omega_{1}$
and $\Omega_{2}$. In the upper panel of Fig. \ref{Fig4} the two
boson energies are $\Omega_{1}=10$ meV and $\Omega_{2}=20$ meV,
while in the lower panel we have $\Omega_{1}=10$ meV and
$\Omega_{2}=30$ meV. The gap values (left axis) and the coupling
constants, $\lambda_{31}$ and $\lambda_{32}$ (right axis) are
plotted as a function of the weight of the low-energy peak, w$_{p}$.
As somehow expected, when the weight of this peak is larger,
(w$_{p}$ = 0.75) the gaps $\Delta_1$ and $\Delta_3$ are larger and
close to the experimental ones but the coupling constants,
$\lambda_{31}$ and $\lambda_{32}$ strongly increase.

In conclusion, we have shown that the newly discovered iron
pnictides very likely represent a case of dominant negative
interband-channel pairing superconductivity where an electron-boson
coupling, such as the electron-SF one, can became a fundamental
ingredient to increase $T_c$ in a multiband strong-coupling picture.
In particular, the present results prove that a simple three-band
model in strong-coupling regime  can reproduce in a quantitative way
the experimental $T_c$ and the energy gaps of the pnictide
superconductors with only two free parameters, $\lambda_{31}$ and
$\lambda_{32}$, provided that the typical energies of the spectral
functions are of the order of 10 meV and the coupling constants are
very high ($\lambda_{31}>4$).

We thank I.I. Mazin and E. Cappelluti for useful discussions.

\end{document}